\documentclass[10pt,journal]{IEEEtran}

\usepackage[T1]{fontenc}
\usepackage[utf8]{inputenc}
\usepackage{amsmath,amssymb}
\usepackage{graphicx}
\usepackage{booktabs}
\usepackage{multirow}
\usepackage{array}
\usepackage{xcolor}
\usepackage{url}
\usepackage{hyperref}
\usepackage{cite}
\usepackage{siunitx}
\usepackage{enumitem}
\usepackage{balance}
\usepackage{dsfont}

\graphicspath{{figures/}}
\hypersetup{colorlinks=true, linkcolor=blue, citecolor=blue, urlcolor=blue}
\begin{document}

\title{Hardware-Aware QAOA for Honeypot Traffic Partitioning on 100+ Qubit IBM Quantum Processors}

\author{Cameron~V.~Cogburn, Casimer~DeCusatis, and~Evan~Spillane
\thanks{C. V. Cogburn is with the Future of Computing Institute, Rensselaer
Polytechnic Institute, Troy, NY 12180 USA (e-mail: cogbuc@rpi.edu).
(Corresponding author: Cameron V. Cogburn.)}
\thanks{C. DeCusatis and E. Spillane are with the School of Computer Science
and Mathematics, Marist University, Poughkeepsie, NY 12601 USA.}
}

\maketitle

\begin{abstract}
Denial-of-service (DoS) and distributed denial-of-service (DDoS) mitigation requires separating malicious traffic from benign traffic while minimizing disruption to legitimate users. 
Prior work proposed mapping honeypot traffic partitioning to a weighted MaxCut problem and solving the resulting graphs with variational quantum algorithms. 
We extend this proof of principle direction
with a reproducible event-level honeypot-to-QUBO pipeline, labeled temporal bipartite benchmark graphs with 16, 32, 66, and 110 event nodes, QAOA executions on IBM quantum hardware, classical heuristic baselines, a noiseless matrix product state reference, and a routing overhead analysis across quantum processor architectures. 
The largest benchmark is a 110-node, 181-edge instance executed on three IBM backends. 
Our results show that a shallow QAOA can execute real traffic partitioning workloads at the utility scale, while backend architecture and routing overhead affect objective quality, security metrics, and observed runtime. 
Because simple classical heuristics can solve the current labeled benchmark graphs, these experiments are not a quantum advantage claim. 
Instead, we deliberately use a fixed, shallow QAOA implementation to enable controlled comparisons across problem sizes and hardware architectures. 
This work establishes a hardware feasibility and architecture benchmark framework, and demonstrates that MaxCut cost, security quality, routing overhead, and runtime must be reported as separate metrics for cybersecurity relevant quantum optimization.
\end{abstract}


\section{Introduction}

DoS and DDoS attacks remain a significant operational threat because defenders must decide which traffic sources or events should be blocked, rate-limited, or preserved. 
A useful mitigation policy is not merely a binary classification; it must isolate attack traffic while preserving benign traffic whenever possible. 
This distinction is important as a mathematically valid partition can cut off malicious nodes, but an ideal mitigation solution should do so while minimizing disruption to legitimate users.

Prior work by DeCusatis et al. proposed representing honeynet traffic as a weighted MaxCut instance and then comparing solving the resulting optimization problem using the variational quantum algorithms QAOA and VQE against classical brute force methods on small DoS/DDoS datasets~\cite{DeCusatis2024QuantumDDoS}. 
In particular, that work introduced the use of non-QML variational quantum algorithms for traffic partitioning and emphasized the distinction between mathematically correct cuts and security optimized cuts. 
It also reported an accuracy--runtime tradeoff between QAOA and VQE on small instances. 
The present paper builds on that direction but focuses on the different question of whether a reproducible QAOA pipeline can execute real traffic partitioning benchmarks on current utility-scale quantum hardware, and how backend architecture affects the resulting objective values, security metrics, and routing overhead cost.

We construct a benchmark set from real honeypot traffic traces.
Each event is represented by one graph node, and each graph node is mapped to one qubit. 
The benchmark graphs are labeled temporal bipartite graphs: attack and benign event labels are used to construct controlled benchmark instances with known class-defined reference cuts. 
This construction is appropriate for a proof of principle hardware study, but is not intended to be a fully autonomous detector. 
In a deployable setting, labels or risk scores would likely be supplied by an upstream intrusion-detection system, firewall rule, anomaly detector, or some other workflow.

The benchmark family contains the honeypot-derived datasets H16, H32, H66, and H110, with the number denoting the number of events, and hence, nodes in the MaxCut graph. 
These instances are used to study shallow QAOA hardware execution, compare QAOA outputs with simple classical baselines, and quantify routing overhead on IBM quantum hardware. 
The single backend scaling study uses \texttt{ibm\_boston} to avoid confounding graph size with backend architecture. 
The H110 architecture study then compares \texttt{ibm\_rensselaer}, \texttt{ibm\_miami}, and \texttt{ibm\_boston} on the same 110-node, 181-edge QAOA workload.

The main contributions of this paper are:
\begin{enumerate}[leftmargin=*]
    \item A reproducible honeypot events to weighted MaxCut workflow for constructing labeled, traffic partitioning benchmark graphs.

    \item Hardware QAOA executions on instances of such benchmark graphs with 16, 32, 66, and 110 event nodes.

    \item A single backend QAOA scaling ladder on \texttt{ibm\_boston} with fixed settings across H16, H32, H66, and H110.

    \item A matched H110 architecture comparison across
    \texttt{ibm\_rensselaer}, \texttt{ibm\_miami}, and \texttt{ibm\_boston}.

    \item Classical baselines using best of 1000 random search, multi-start greedy local search, and simulated annealing, which establish that the benchmark graphs studied are classically easy under simple heuristics, hence, bound the interpretation of the quantum results.

    \item A noiseless matrix product state QAOA reference that attributes most of the cost ratio gap to the shallow ansatz and limited optimizer budget rather than to hardware noise.

    \item A routing overhead analysis quantifying transpiled depth, two-qubit gate counts, estimated SWAP-equivalent overhead, and two-qubit/logical interaction ratios for the QAOA circuits.

    \item A security metric framework that separates MaxCut cost ratio from attack recall, benign preservation, quarantine precision, and balanced security score.
\end{enumerate}

\noindent\textbf{Scope.}
This paper is a proof of principle quantum engineering benchmark. 
It does not claim quantum advantage over classical heuristics, nor does it claim to solve the upstream problem of assigning attack labels to previously unseen traffic.
Instead, it establishes a reproducible experimental framework for executing QAOA workloads built from real traffic traces on quantum hardware, comparing hardware architectures, auditing routing overhead, and evaluating optimization and security metrics separately. 
Studies of label-free and label-assisted graph construction, security-aware QUBO objectives, systematic QAOA--VQE comparisons, and energy-aware cost of solution benchmarking are left for future work.

\section{Related Work}
Quantum and quantum-inspired techniques have been explored for cybersecurity, including quantum machine learning classifiers and quantum-inspired optimization methods \cite{Said2023QCandMLDDoS,Rivas2024QuantumEnhancedRepDDoS,Gong2019QuantumGeneticDDoS}. 
The most direct precursor to this work is by DeCusatis et al., which formulated honeynet traffic partitioning as a MaxCut problem and compared QAOA and VQE on small attack graphs \cite{DeCusatis2024QuantumDDoS}. 
That work reported strong small scale feasibility results, compared QAOA and VQE under noise and noiseless conditions, and extrapolated execution time behavior.

Our work differs in four ways. 
First, we rebuild the formulation as a reproducible and scalable workflow using event-level data rather than relying on the original, bespoke implementation. 
Second, we scale the hardware experiments to a 110-event/110-qubit benchmark instance. 
Third, we compare multiple IBM hardware architectures rather than reporting a single small device result. 
Fourth, we add classical heuristics and routing metrics to clarify what is and is not being demonstrated.

QAOA was introduced by Farhi et al. as a variational algorithm for combinatorial optimization problems \cite{farhi2014qaoa}. 
VQE is a broader variational framework originally developed for eigenvalue problems \cite{peruzzo2014vqe}. 
MaxCut is a canonical NP-hard graph-partitioning problem with an extensive classical approximation literature, including semidefinite-programming approximations \cite{Hastad2001OptimalInapprox, Goemans1995Maxcut}. 
Here, we utilize MaxCut as a benchmark formulation for quantum hardware execution and security oriented partition evaluation.

Prior work has executed QAOA on superconducting hardware at scales comparable to ours. 
Harrigan et al. reported QAOA on up to 23 qubits on the
Google Sycamore processor, including non-native MaxCut problems requiring significant compilation, and observed that performance on non-native graphs degrades with problem size~\cite{Harrigan2021QAOA}. 
Weidenfeller et al. analyzed SWAP strategies for QAOA on linear, grid, and heavy-hex coupling maps, and provided a Qiskit Runtime QAOA program targeted at IBM hardware~\cite{Weidenfeller2022Scaling}. 
More recently, Pelofske et al. executed $p=1$--$5$ QAOA on ensembles of higher order Ising instances on IBM heavy-hex devices with up to 127 qubits, using matrix product state simulation to provide noise-free baselines and reporting that the best quantum processors find lower energy solutions up to $p\approx2$--$3$ before noise dominates~\cite{Pelofske2024Scaling}. 

At a comparable scale, Wang et al. executed a depth-independent linear-chain QAOA ansatz on IBM superconducting processors for non-hardware-native random-regular MaxCut instances with up to 100 vertices, restructuring the circuit so that depth is independent of problem size~\cite{Wang2025LinearChainQAOA}. 
Sachdeva et al. paired an error suppressed QAOA pipeline with a 127-qubit IBM processor to solve random-regular MaxCut instances whose topology is not matched to device connectivity~\cite{Sachdeva2024QAOA156}.

Our work differs from these studies in two ways. 
First, the benchmark workload is constructed from real cybersecurity event traces rather than synthetic Ising or random regular graphs, which produces a partially structured graph family relevant to a deployable security application. 
Second, we directly compare Heron r3, Nighthawk r1, and Eagle IBM architectures on a matched 110-node logical workload, isolating backend effects on objective quality, security quality, routing overhead, and observed runtime.

\subsection*{Scope of Variational Quantum Algorithms}
Prior work by DeCusatis et al. compared QAOA and VQE for small traffic partitioning MaxCut instances and reported that VQE could be more accurate in certain noisy cases, but required substantially longer execution times \cite{DeCusatis2024QuantumDDoS}. 
In the present work, we do not attempt to reproduce a full QAOA vs.~VQE comparison. 
Rather, we focus on QAOA as the primary variational optimizer because the goal is to evaluate large honeypot graphs on current quantum hardware \cite{Roca2021VQAs}.
This choice keeps the experimental scope here focused on hardware architecture and routing. 
We defer to a future work for large scale VQE benchmarking and optimization.

\section{Problem Formulation}
Here we provide details for the formulation of the problem. 

\subsection{Weighted MaxCut}
Let \(G=(V,E,w)\) be a weighted undirected graph with node set \(V=\{1,\ldots,n\}\), edge set \(E \subseteq \binom{V}{2}\), and nonnegative edge weights \(w_{ij} \geq 0\) for \((i,j)\in E\). 
A binary partition can be encoded via
\begin{equation}
    x_i \in \{0,1\}, \qquad i \in V.
\end{equation}
The weighted MaxCut objective is
\begin{align}
    C(x) &= \sum_{(i,j)\in E} w_{ij} \left[x_i(1-x_j) + x_j(1-x_i)\right] \nonumber \\
    &= \sum_{(i,j)\in E} w_{ij} (x_i+x_j-2x_ix_j) ,
    \label{eq:maxcut-binary}
\end{align}
or equivalently,
\begin{equation}
    C(x) = \sum_{(i,j)\in E} w_{ij}\, \mathds{1} [x_i \neq x_j].
\end{equation}
As can be seen in (\ref{eq:maxcut-binary}), since each edge contributes a quadratic term, this is a quadratic unconstrained binary optimization (QUBO) objective.
For nonnegative weights, maximizing \(C(x)\) means finding the partition that cuts as much edge weight as possible.

\subsection{Temporal Bipartite Graphs}
In this work we focus on graphs constructed from labeled honeypot traffic events, which are mathematically temporal bipartite graphs.
By labeled we mean attack events and benign events are known from the honeypot trace. 
Edges are then created between attack events and nearby benign events with weights determined by temporal proximity and a fixed separation weight:
\begin{equation}
w_{ij}=W_{\mathrm{sep}}\exp\left(-\frac{|t_i-t_j|}{\tau}\right),
\label{eq:weights}
\end{equation}
where
\begin{equation}
    W_{\mathrm{sep}}=5.0,\qquad \tau=2.0.
\end{equation}
The result of this labeled construction is a known reference cut: since every edge connects an attack event to a benign event, separating all of these cuts all of the edges. 
We therefore use the resulting cut value 
\begin{equation}
C_{\mathrm{ref}}=\sum_{(i,j)\in E} w_{ij}
\label{eq:C_ref}
\end{equation}
as the benchmark objective for evaluating the classical baselines and quantum algorithms.

The purpose of this construction is to provide a controlled benchmark for testing whether QAOA can execute traffic partitioning MaxCut instances on current superconducting quantum hardware, and how different architectures affect performance.

\subsection{Cost Ratio}
Following prior work \cite{DeCusatis2024QuantumDDoS}, we define the MaxCut cost ratio
\begin{equation}
    \rho_C = \frac{C(x)}{C_{\mathrm{ref}}}, \quad C_{\mathrm{ref}}=C(x^*)
\end{equation}
where \(x^*\) is the reference oracle solution for the labeled benchmark graph. 
The formula for \(C_{\mathrm{ref}}\) was given in (\ref{eq:C_ref}).
For the labeled temporal bipartite benchmark graphs used here, the class-defined reference cut cuts every edge and is therefore the MaxCut optimum for this constructed graph model. 
For other graph constructions, such as label-free or signed-weight graphs, the reference value would need to be computed separately for small instances or replaced by another benchmark value.

\subsection{Security Metrics}

Distinguishing between a mathematically correct and ideal security solution is essential as a MaxCut solution may be mathematically correct without being an ideal security solution for several reasons \cite{DeCusatis2024QuantumDDoS}. 
First, MaxCut returns an unordered bipartition, so the quarantine and benign sides must be assigned after the fact. 
Second, a cut may isolate attack traffic while also quarantining benign traffic. 
Third, when the graph has multiple connected components, different component orientations can preserve the same MaxCut objective while changing the global security interpretation. 
Finally, the MaxCut objective only encodes the edge weights supplied to it, therefore unless benign preservation or false positive costs are represented in the graph, a high objective value does not necessarily imply an ideal mitigation policy.

For all of these reasons, we evaluate each partition not only with optimization metrics, but with the following security metrics.

Let \(y_i\in\{\mathrm{attack},\mathrm{benign}\}\) denote the class label of event \(i\), and let \(x_i\in\{0,1\}\) denote the partition bit returned by the solver. 
The MaxCut objective is invariant under a global bit flip with \(x\) and \(1-x\) defining the same cut. 
To be able to evaluate a partition as a mitigation
decision, we first orient the partition by choosing the side containing the majority of attack events as the quarantine side:
\begin{equation}
    s_Q(x)
    =
    \operatorname*{arg\,max}_{s\in\{0,1\}}
    \left|\{i: y_i=\mathrm{attack},\ x_i=s\}\right|.
\end{equation}
The corresponding quarantine set is
\begin{equation}
    Q(x) = \{i: x_i=s_Q(x)\}.
\end{equation}
Let
\begin{equation}
    A = \{i: y_i=\mathrm{attack}\}, \qquad
    B = \{i: y_i=\mathrm{benign}\}.
\end{equation}
We then compute
\begin{align}
    \text{Attack Recall}
    &= \frac{|A\cap Q|}{|A|}, \\
    \text{Benign Preservation}
    &= \frac{|B\setminus Q|}{|B|}, \\
    \text{Quarantine Precision}
    &= \frac{|A\cap Q|}{|Q|}, \\
    \text{Balanced Security} 
    =  &\frac{1}{2}
    \left(
    \text{Attack Recall}
    +
    \text{Benign Preservation} 
    \right)
\end{align}

The balanced security score is not intended to replace recall or precision.
It is a simple and compact plotting metric that rewards partitions which simultaneously quarantine attack events and preserve benign events.

Note the orientation step is used only for retrospective evaluation on the labeled benchmark data in this work. 
In a deployable system the quarantine side would be chosen using an upstream detector, honeypot rule, analyst label, etc.

\section{Experimental Design and Implementation}

\begin{figure*}[t]
\centering
\includegraphics[width=0.95\textwidth]{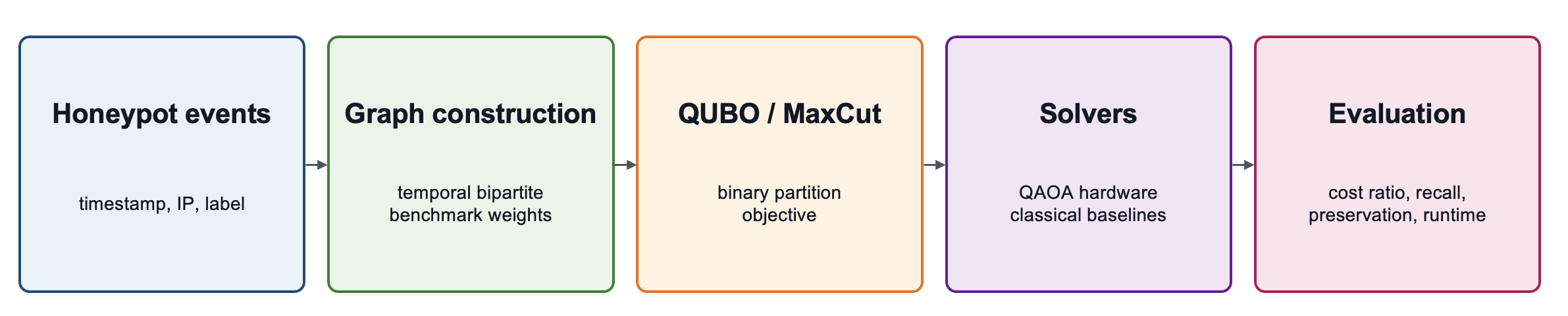}
\caption{Workflow used in this study. Labeled honeypot event traces are converted into temporal bipartite MaxCut benchmark graphs, solved with QAOA hardware runs and classical baselines, and evaluated with both optimization and security metrics.
}
\label{fig:pipeline}
\end{figure*}

Figure~\ref{fig:pipeline} summarizes the overall workflow, from honeypot event traces through graph construction, solving, and evaluation. 
The remainder of this section details each stage.

\subsection{Datasets}

The dataset consists of four honeypot-derived benchmark instances, H16, H32, H66, and H110, corresponding to real honeypot-derived graphs with 16, 32, 66, and 110 events, respectively.

\begin{table*}[t]
\centering
\caption{Honeypot-derived benchmark datasets and graph sizes.} 
\label{tab:dataset-summary}
\resizebox{\textwidth}{!}{
\begin{tabular}{lrrrrr}
\toprule
Dataset & Event nodes / qubits & Attack events & Benign events & Graph edges & Class-defined reference objective \\
\midrule
H16 & 16 & 12 & 4 & 24 & 74.290 \\
H32 & 32 & 20 & 12 & 43 & 145.093 \\
H66 & 66 & 43 & 23 & 92 & 315.326 \\
H110 & 110 & 90 & 20 & 181 & 677.633 \\
\bottomrule
\end{tabular}

}
\end{table*}

\subsection{QAOA Hardware Runs}

\label{sec:QAOA}
We intentionally use a generic, shallow QAOA configuration and avoid backend specific algorithm tuning. 
All hardware runs use \(p=1\), COBYLA, 1024 shots, and 15 optimizer evaluations. 
H16, H32, and H66 use five seeds (124--128); H110 uses six seeds (124--129) in both the scaling study and the
architecture comparison of Section~\ref{sec:results-arch}. 
Exact seed counts are reported in the corresponding tables. 
This choice is made to isolate problem size, backend architecture, and routing effects. 
More aggressive QAOA tuning, including larger \(p\), alternative optimizers, backend specific layouts, and transpiler optimization level sweeps, is left for future work. 
All circuits, transpilation, and Runtime executions use Qiskit~\cite{Javadiabhari2024Qiskit}.

Using a depth \(p=1\) QAOA circuit, we alternate a problem-dependent cost layer and a mixer layer. 
The cost layer applies two-qubit phase interactions for every graph edge, so the number and structure of graph edges directly affect the transpiled circuit depth and routing overhead. 
The mixer layer applies single-qubit rotations. 
We optimize the QAOA angles with the COBYLA optimizer \cite{Powell1994} and sample the resulting circuit with \(1024\) shots per evaluation.

For the 32-, 66-, and 110-qubit hardware runs, we use a manual Qiskit Runtime SamplerV2 optimization loop. 
At each optimizer evaluation, the parameterized QAOA circuit is sampled with 1024 shots and the sampled mean cut value is passed to COBYLA. 
The reported bitstring is the best finite shot sample by objective value, rather than the most probable bitstring.
This choice is important for large shot, high qubit runs, where many bitstrings can appear only once.

\subsection{Classical Baselines}
\label{sec:classical-baselines}

We compare the QAOA hardware results against three classical baselines:
\begin{enumerate}[leftmargin=*]
    \item best of 1000 random partitions;
    \item multi-start greedy local search with 100 random starts;
    \item simulated annealing with 100 reads and 1000 steps per read.
\end{enumerate}

These baselines serve two purposes. First, they provide a sanity check: if a quantum hardware run cannot outperform random partitions, then its output is not useful even as a proof of principle. 
Second, they calibrate the difficulty of the benchmark graphs themselves, which fixes the scope of the comparison. 
The results of this comparison are reported in Section~\ref{sec:results-classical}.

\subsection{Hardware Backends and Routing Analysis}

We use IBM superconducting quantum processors for two related but distinct experiments. 
First, we perform a scaling study on \texttt{ibm\_boston}, using H16, H32, H66, and H110 with identical QAOA settings. 
Second, we perform an H110 architecture comparison across \texttt{ibm\_rensselaer}, \texttt{ibm\_miami}, and \texttt{ibm\_boston}.

The fixed backend scaling study uses \texttt{ibm\_boston} to avoid confounding graph size with backend architecture. 
Boston was selected after the H110 architecture screen because it produced the strongest mean cost ratio among the three tested backends and beat the best of 1000 random baseline in more of its H110 seeds than either other backend (Table~\ref{tab:h110-architecture}). 

The architecture comparison uses H110 because it is the largest instance in the benchmark dataset and therefore the most sensitive to routing overhead. 
The three backends represent distinct access and architecture regimes. 
The \texttt{ibm\_rensselaer} backend is a 127-qubit Eagle processor with a heavy-hex coupling map and serves as the institutional baseline available to this project. 
The \texttt{ibm\_miami} backend is a Nighthawk r1 processor with a grid topology and increased connectivity, while \texttt{ibm\_boston} is a Heron r3 processor. 
These choices allow us to compare how backend topology, routing overhead, and hardware performance affect the same 110-node QAOA workload.

For the circuit routing analysis, the QAOA circuits are transpiled to each backend at optimization level 1 without submitting a QPU job. 
These routing metrics report transpiled depth, two-qubit gate count, estimated SWAP-equivalent overhead, and the ratio of routed two-qubit gates to logical graph interactions.

\section{Results}

\subsection{Boston QAOA Scaling Across H16--H110}

We first fix the quantum backend to \texttt{ibm\_boston} and evaluate QAOA on the four frozen honeypot benchmark graphs H16, H32, H66, and H110.
This isolates graph size from backend architecture. 
Table~\ref{tab:boston-qaoa-scaling} summarizes the resulting cost ratios, security metrics, random-baseline success rates, and observed wall-clock time per optimizer evaluation.

\begin{figure}[t]
\centering
\includegraphics[width=\columnwidth]{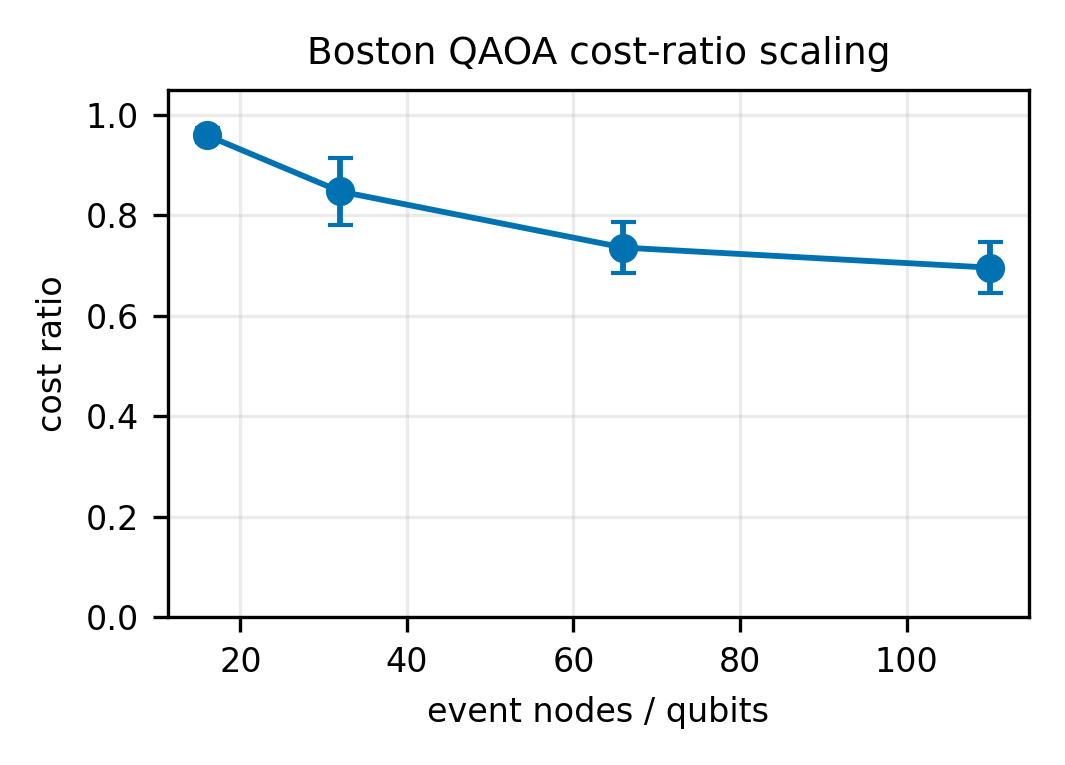}
\caption{Boston QAOA cost ratio scaling over H16, H32, H66, and H110. 
Error bars indicate sample standard deviation over five seeds (124--128) for H16--H66 and six seeds (124--129) for H110.}
\label{fig:boston-cost-scaling}
\end{figure}

Figure~\ref{fig:boston-cost-scaling} shows the cost ratio trend across the four graph sizes. The mean cost ratio decreases from H16 to H110, which is expected for a fixed shallow QAOA circuit and fixed optimizer budget. 
The hardware runs remain nontrivial: QAOA beats the best of 1000 random objective baseline in 20 of the 21 Boston scaling runs, missing only one of the six H110 seeds (Table~\ref{tab:boston-qaoa-scaling}).

\begin{figure}[t]
\centering
\includegraphics[width=\columnwidth]{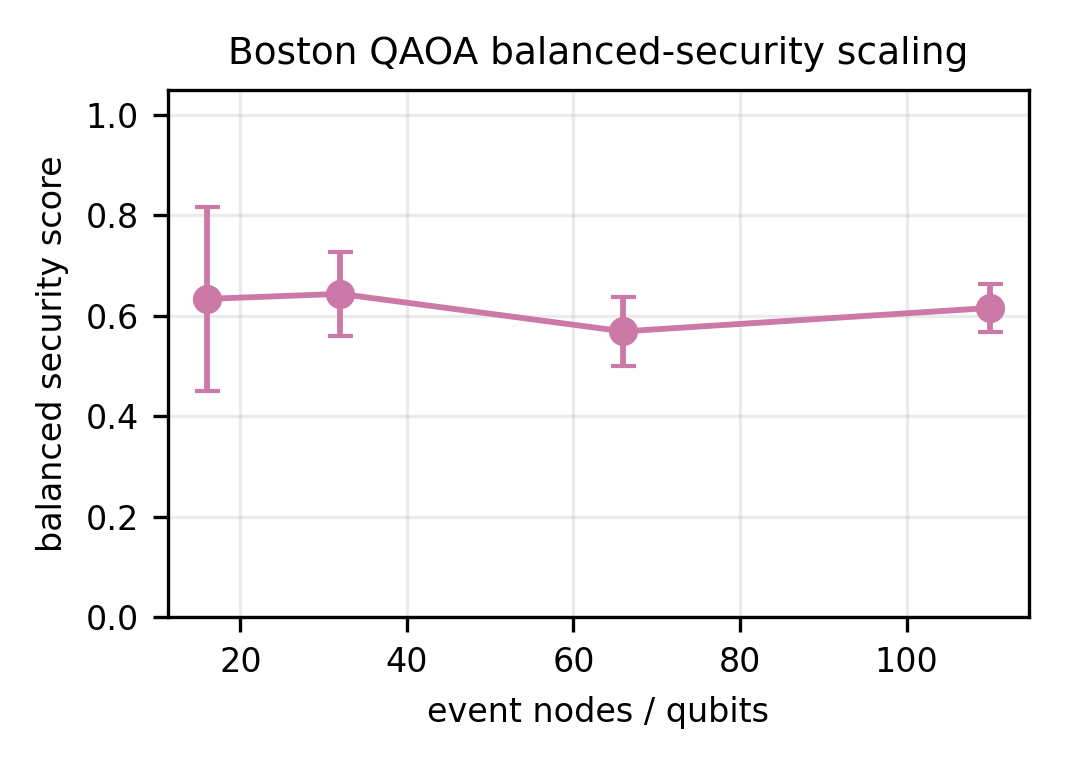}
\caption{Balanced security score for the Boston QAOA scaling ladder. 
This metric averages attack recall and benign preservation.}
\label{fig:boston-security-scaling}
\end{figure}

Figure~\ref{fig:boston-security-scaling} shows the corresponding balanced security score. 
Unlike the cost ratio, the balanced security score is not monotonic in graph size. 
This indicates that MaxCut objective value and mitigation quality are related but distinct quantities, motivating the separate reporting of attack recall, benign preservation, quarantine precision, and
balanced security throughout this work.

\begin{table*}[t]
\centering
\caption{
Boston QAOA scaling results. Values are mean \(\pm\) sample standard deviation.
H16, H32, and H66 use five seeds, 124--128 while H110 uses six seeds, 124--129, matching the Boston H110 aggregate used in the architecture comparison.
}
\label{tab:boston-qaoa-scaling}
\resizebox{\textwidth}{!}{
\begin{tabular}{lrrrllllll}
\toprule
Dataset & Nodes & Edges & Runs & Cost ratio & Attack recall & Benign preservation & Balanced security & Beats random best & Time/eval (s) \\
\midrule
H16 & 16 & 24 & 5 & 0.959 $\pm$ 0.014 & 0.667 $\pm$ 0.144 & 0.600 $\pm$ 0.224 & 0.633 $\pm$ 0.183 & 5/5 & 129.6 $\pm$ 110.3 \\
H32 & 32 & 43 & 5 & 0.847 $\pm$ 0.067 & 0.670 $\pm$ 0.076 & 0.617 $\pm$ 0.095 & 0.643 $\pm$ 0.083 & 5/5 & 18.1 $\pm$ 16.2 \\
H66 & 66 & 92 & 5 & 0.735 $\pm$ 0.051 & 0.572 $\pm$ 0.073 & 0.565 $\pm$ 0.081 & 0.569 $\pm$ 0.068 & 5/5 & 107.8 $\pm$ 90.9 \\
H110 & 110 & 181 & 6 & 0.696 $\pm$ 0.051 & 0.656 $\pm$ 0.048 & 0.575 $\pm$ 0.069 & 0.615 $\pm$ 0.048 & 5/6 & 295.3 $\pm$ 412.5 \\
\bottomrule
\end{tabular}

}
\end{table*}

\subsection{Noiseless Finite-Shot MPS Baseline}
\label{sec:results-mps}

To separate the effect of the shallow \(p=1\) ansatz and the limited optimizer budget from the effect of hardware noise and routing, we repeated the Boston QAOA workflow on a noiseless Aer matrix product state (MPS) simulator. 
The simulator runs reuse the hardware configuration, COBYLA with 1024 shots and 15 optimizer evaluations, using five seeds (124--128) for H16, H32, and H66, and six seeds (124--129) for H110.
The matrix product state simulation of H110 was costly at roughly \(2.7\times 10^{4}\) seconds per seed on our local configuration\footnote{Apple MacBook Pro M4 Max with 64 GB of memory.}. We therefore treat the H110 simulator timings only as a record of simulation cost rather than as a hardware comparison metric.

\begin{figure*}[t]
\centering
\includegraphics[width=\textwidth]{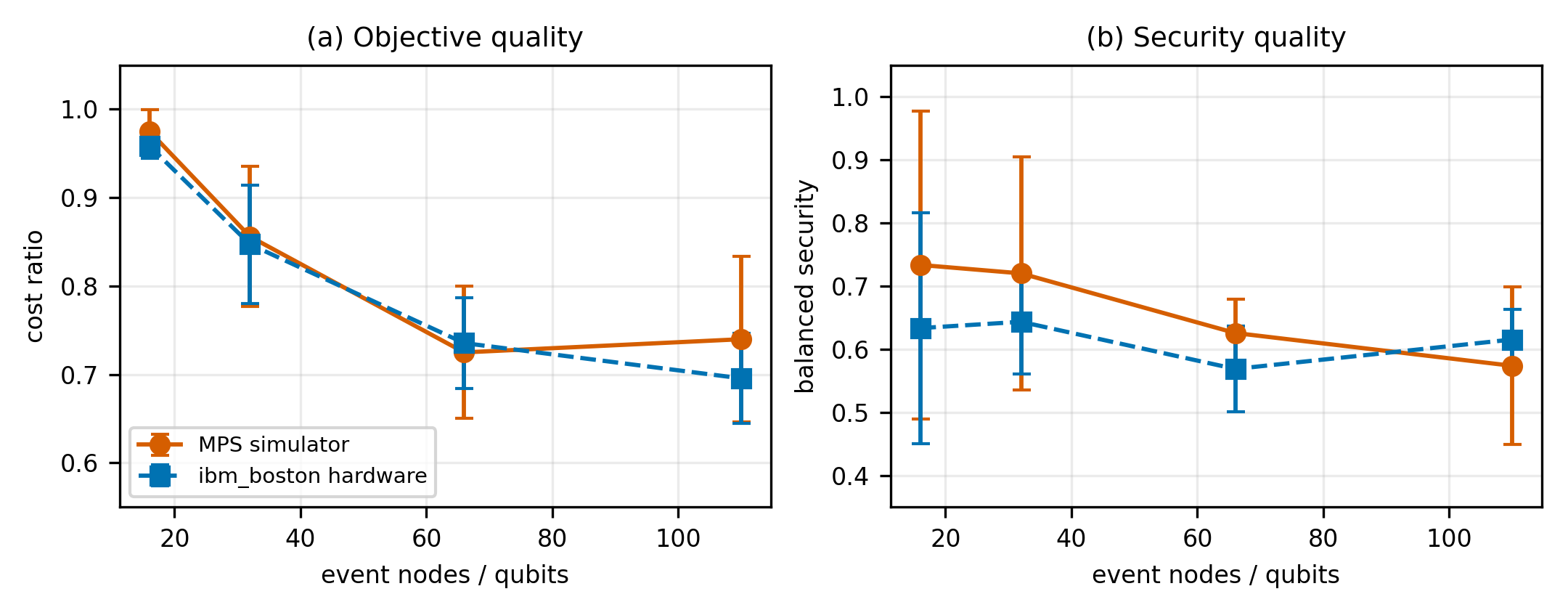}
\caption{Noiseless finite-shot MPS QAOA baseline compared with
\texttt{ibm\_boston} hardware QAOA. Both workflows use \(p=1\),
COBYLA, 1024 shots, and 15 optimizer evaluations. Markers show means
and error bars show sample standard deviations. 
The MPS baseline and the Boston hardware both use
five seeds (124--128) for H16, H32, and H66, while H110 included an additional sixth seed (129).}
\label{fig:mps-vs-hardware}
\end{figure*}

Table~\ref{tab:mps-baseline} reports the simulator results, and  Fig.~\ref{fig:mps-vs-hardware} overlays them on the Boston hardware data. 
The two cost ratio curves track each other within the sample standard deviation across all four sizes, and the noiseless simulator does not substantially outperform the hardware. 
The key observation is that even without noise the simulator plateaus well below the class-defined reference objective, reaching only about 0.74 at H110. 
The roughly 0.25 shortfall is therefore largely algorithmic, attributable to the shallow ansatz, the finite shot count, the limited optimizer budget, and the stochastic best-sample selection, while the residual difference attributable to hardware noise is small and comparable to the seed-to-seed scatter (about 0.04 in mean cost ratio at H110). 
The simulator thus provides a useful algorithm-only reference for the largest workload.

\begin{table*}[t]
\centering
\caption{Noiseless finite-shot MPS QAOA baseline versus
\texttt{ibm\_boston} hardware QAOA. Both use the same QAOA
configuration, \(p=1\), COBYLA, 1024 shots, and 15 optimizer
evaluations. The MPS baseline and the Boston hardware columns both use
five seeds (124--128) for H16, H32, and H66, while H110 included an additional sixth seed (129).}
\label{tab:mps-baseline}
\resizebox{\textwidth}{!}{
\begin{tabular}{lrrrllrlll}
\toprule
Dataset & Nodes & Edges & MPS seeds & MPS cost ratio & Boston hardware seeds & Boston hardware cost ratio & MPS balanced security & Boston hardware balanced security & MPS time/eval (s) \\
\midrule
H16 & 16 & 24 & 5 & 0.975 $\pm$ 0.024 & 5 & 0.959 $\pm$ 0.014 & 0.733 $\pm$ 0.244 & 0.633 $\pm$ 0.183 & 0.03 $\pm$ 0.02 \\
H32 & 32 & 43 & 5 & 0.856 $\pm$ 0.079 & 5 & 0.847 $\pm$ 0.067 & 0.720 $\pm$ 0.184 & 0.643 $\pm$ 0.083 & 0.04 $\pm$ 0.01 \\
H66 & 66 & 92 & 5 & 0.725 $\pm$ 0.075 & 5 & 0.735 $\pm$ 0.051 & 0.626 $\pm$ 0.053 & 0.569 $\pm$ 0.068 & 0.09 $\pm$ 0.00 \\
H110 & 110 & 181 & 6 & 0.740 $\pm$ 0.094 & 6 & 0.696 $\pm$ 0.051 & 0.574 $\pm$ 0.125 & 0.615 $\pm$ 0.048 & 1809.82 $\pm$ 30.48 \\
\bottomrule
\end{tabular}

}
\end{table*}

\subsection{Comparison with Classical Baselines}
\label{sec:results-classical}

\begin{figure}[t]
\centering
\includegraphics[width=\columnwidth]{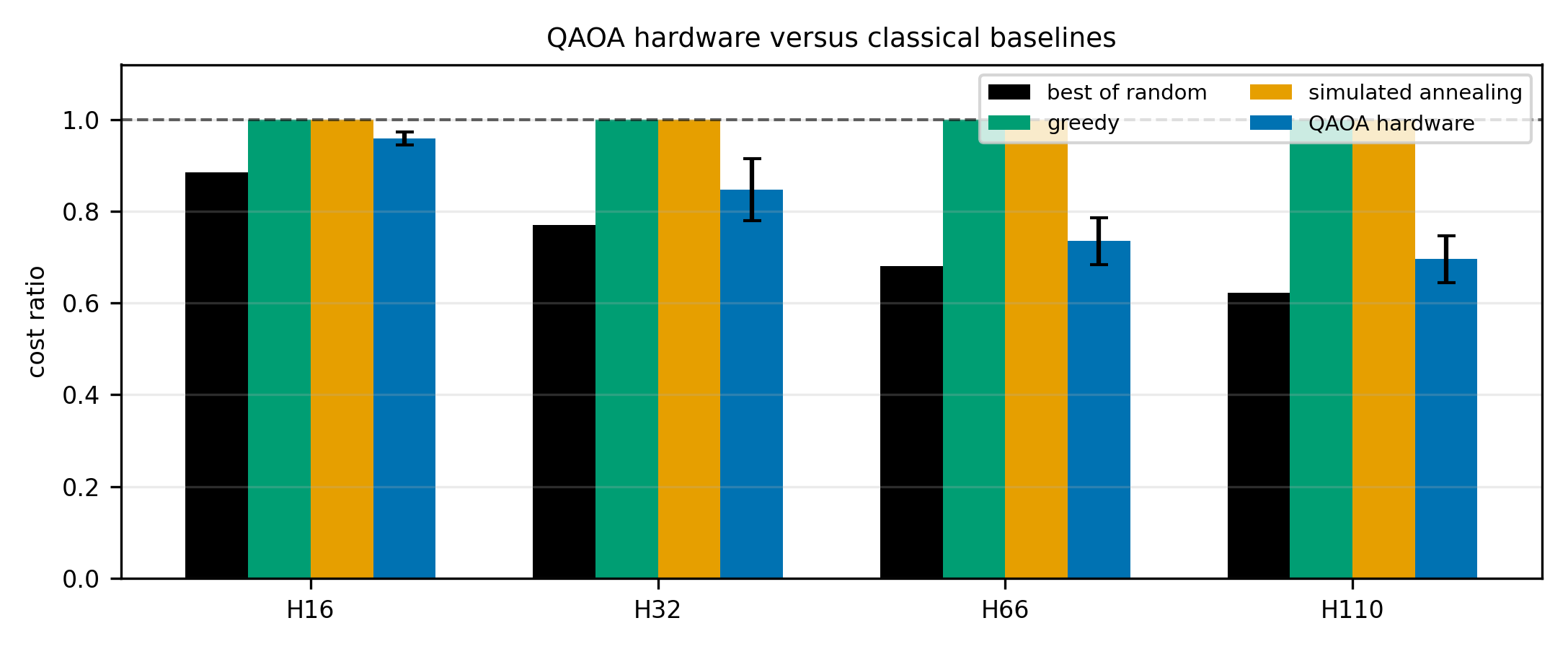}
\caption{QAOA hardware cost ratios compared with best of 1000 random, greedy local search, and simulated annealing. 
Greedy local search and simulated annealing reach the class-defined reference objective on all four labeled temporal bipartite benchmark graphs.}
\label{fig:qaoa-vs-classical-cost}
\end{figure}

Table~\ref{tab:qaoa-vs-classical} and Fig.~\ref{fig:qaoa-vs-classical-cost} compare the Boston QAOA hardware results against the three classical baselines of Section~\ref{sec:classical-baselines}. 
Greedy local search and simulated annealing both reach the class-defined reference MaxCut objective on H16, H32, H66, and H110, while QAOA exceeds the best of 1000 random baseline in 20 of the 21 Boston runs without reaching the reference.

These numbers fix the scope of the paper rather than diminish it. Because the labeled temporal bipartite graphs are classically easy, the experiments are a controlled test of hardware feasibility, not a claim of quantum advantage. 
What the quantum runs establish is execution at 100+ qubits, architecture-dependent behavior, measurable routing overhead, and the separation between MaxCut objective value and security specific mitigation quality.

\begin{table*}[t]
\centering
\caption{QAOA hardware results compared with classical baselines. 
Cost ratios are relative to the class-defined reference objective.}
\label{tab:qaoa-vs-classical}
\resizebox{\textwidth}{!}{
\begin{tabular}{lrlllllllll}
\toprule
Dataset & Nodes & Random best cost ratio & Random best balanced security & Greedy cost ratio & Greedy balanced security & SA cost ratio & SA balanced security & QAOA mean cost ratio & QAOA mean balanced security & QAOA beats random best \\
\midrule
H16 & 16 & 0.886 & 0.500 & 1.000 & 1.000 & 1.000 & 1.000 & 0.959 $\pm$ 0.014 & 0.633 $\pm$ 0.183 & 5/5 \\
H32 & 32 & 0.770 & 0.525 & 1.000 & 0.592 & 1.000 & 0.817 & 0.847 $\pm$ 0.067 & 0.643 $\pm$ 0.083 & 5/5 \\
H66 & 66 & 0.680 & 0.588 & 1.000 & 0.562 & 1.000 & 0.788 & 0.735 $\pm$ 0.051 & 0.569 $\pm$ 0.068 & 5/5 \\
H110 & 110 & 0.623 & 0.503 & 1.000 & 0.819 & 1.000 & 0.808 & 0.696 $\pm$ 0.051 & 0.615 $\pm$ 0.048 & 5/6 \\
\bottomrule
\end{tabular}

}
\end{table*}

\subsection{H110 Architecture Comparison}
\label{sec:results-arch}

We next fix the graph to H110 and compare three IBM quantum backends: \texttt{ibm\_rensselaer}, \texttt{ibm\_miami}, and \texttt{ibm\_boston}.
The H110 graph is the largest benchmark instance in the study, with 110 event nodes and 181 logical graph edges, making it the most sensitive to backend routing and hardware effects. 
All architecture comparison runs use QAOA with the same optimizer budget and shot count as the Boston scaling study.

\begin{figure}[t]
\centering
\includegraphics[width=\columnwidth]{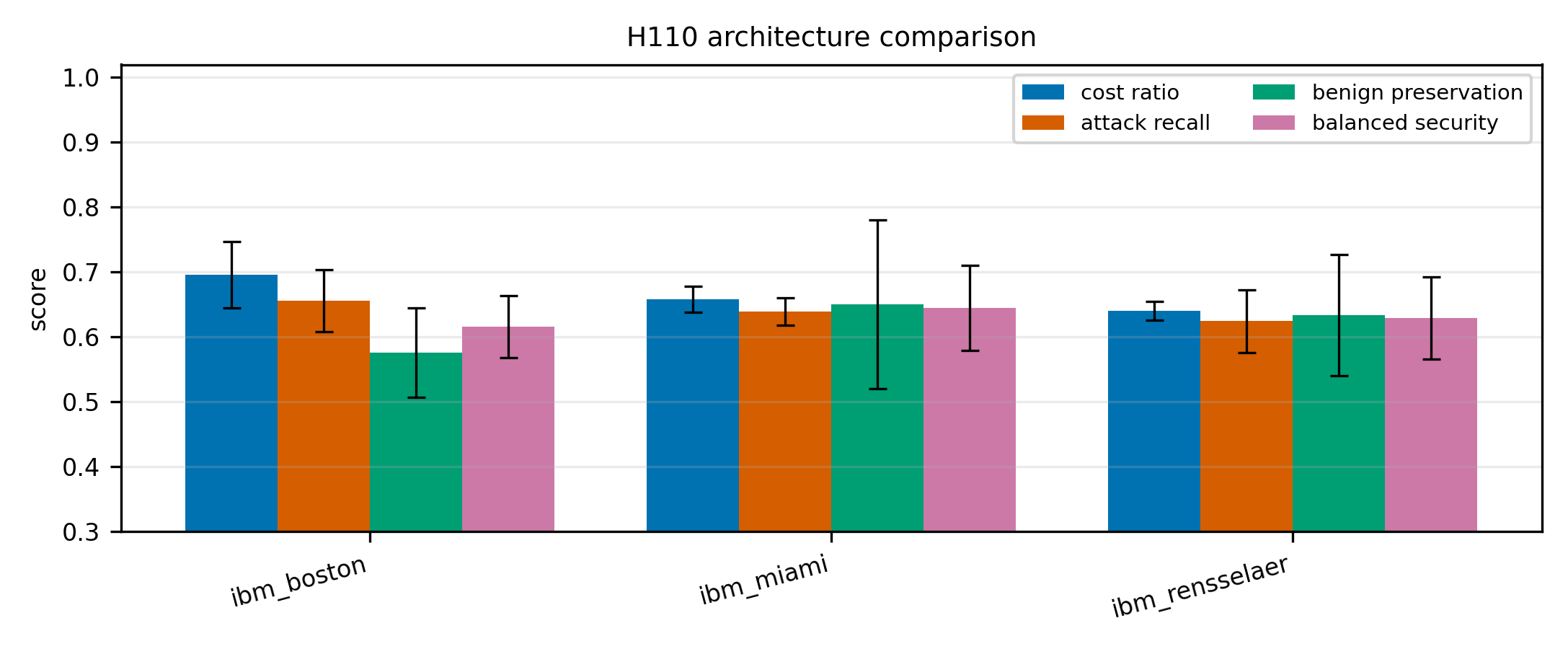}
\caption{H110 architecture comparison. 
\texttt{ibm\_boston} achieves the largest mean cost ratio, while \texttt{ibm\_miami} achieves the largest mean balanced security score.}
\label{fig:h110-architecture-quality}
\end{figure}

\begin{figure}[t]
\centering
\includegraphics[width=\columnwidth]{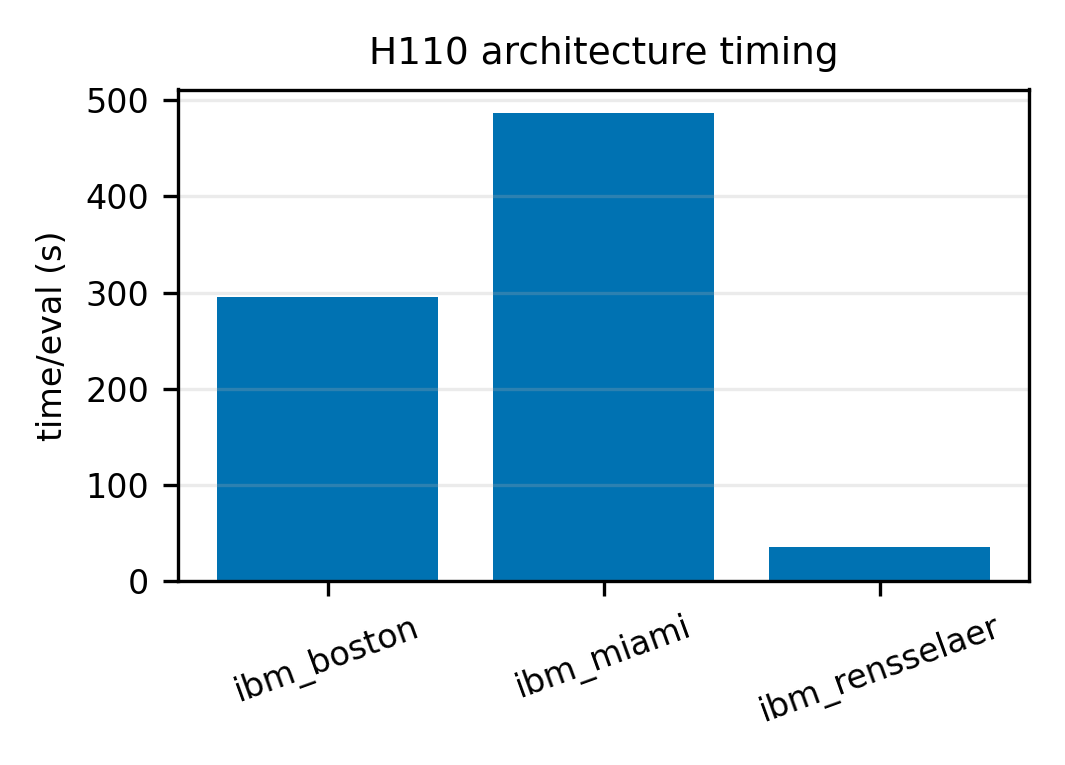}
\caption{Observed wall-clock time per optimizer evaluation for the H110 architecture comparison. 
These measurements include IBM Runtime session behavior, job
submission, result retrieval, and cloud execution variability, and should not be interpreted as isolated QPU execution time.}
\label{fig:h110-architecture-timing}
\end{figure}

Table~\ref{tab:h110-architecture} reports the aggregate H110 results by backend. 
Figure~\ref{fig:h110-architecture-quality} compares mean cost ratio, attack recall, benign preservation, and balanced security. Figure~\ref{fig:h110-architecture-timing} compares the observed wall-clock time per optimizer evaluation.

Across the H110 runs, \texttt{ibm\_boston} achieves the largest mean cost ratio and beats the best of 1000 random objective baseline in five of its six seeds, more often than either other backend. 
\texttt{ibm\_miami} achieves the highest mean balanced security score, and \texttt{ibm\_rensselaer}, the institutional baseline, has the lowest mean cost ratio of the three. 
Mean wall-clock time per optimizer evaluation varies widely across backends but does not track objective quality or routing overhead (Section~\ref{sec:runtime-caveats}), so we do not interpret it as a backend performance metric. 
Overall, backend choice affects objective quality and security quality even when the graph, algorithm, optimizer budget, and shot count are held fixed.

\begin{table*}[t]
\centering
\caption{Matched H110 QAOA architecture comparison across IBM backends.}
\label{tab:h110-architecture}
\resizebox{\textwidth}{!}{
\begin{tabular}{lrlllllll}
\toprule
Backend & Runs & Mean cost ratio & Std cost ratio & Mean attack recall & Mean benign preservation & Mean balanced security & Beats random best & Mean time/eval (s) \\
\midrule
\texttt{ibm\_boston} & 6 & 0.696 & 0.051 & 0.656 & 0.575 & 0.615 & 5/6 & 295.3 \\
\texttt{ibm\_miami} & 6 & 0.657 & 0.020 & 0.639 & 0.650 & 0.644 & 4/6 & 486.5 \\
\texttt{ibm\_rensselaer} & 6 & 0.640 & 0.015 & 0.624 & 0.633 & 0.629 & 3/6 & 36.0 \\
\bottomrule
\end{tabular}

}
\end{table*}

\subsection{Circuit Routing and Transpilation Overhead}

QAOA cost layers require two-qubit interactions corresponding to the logical edges of the MaxCut graph. 
On real hardware, those logical interactions must be mapped onto the backend coupling graph. 
Routing can therefore add substantial two-qubit overhead, especially when logical graph edges do not align with physical qubit connectivity.

\begin{figure}[t]
\centering
\includegraphics[width=\columnwidth]{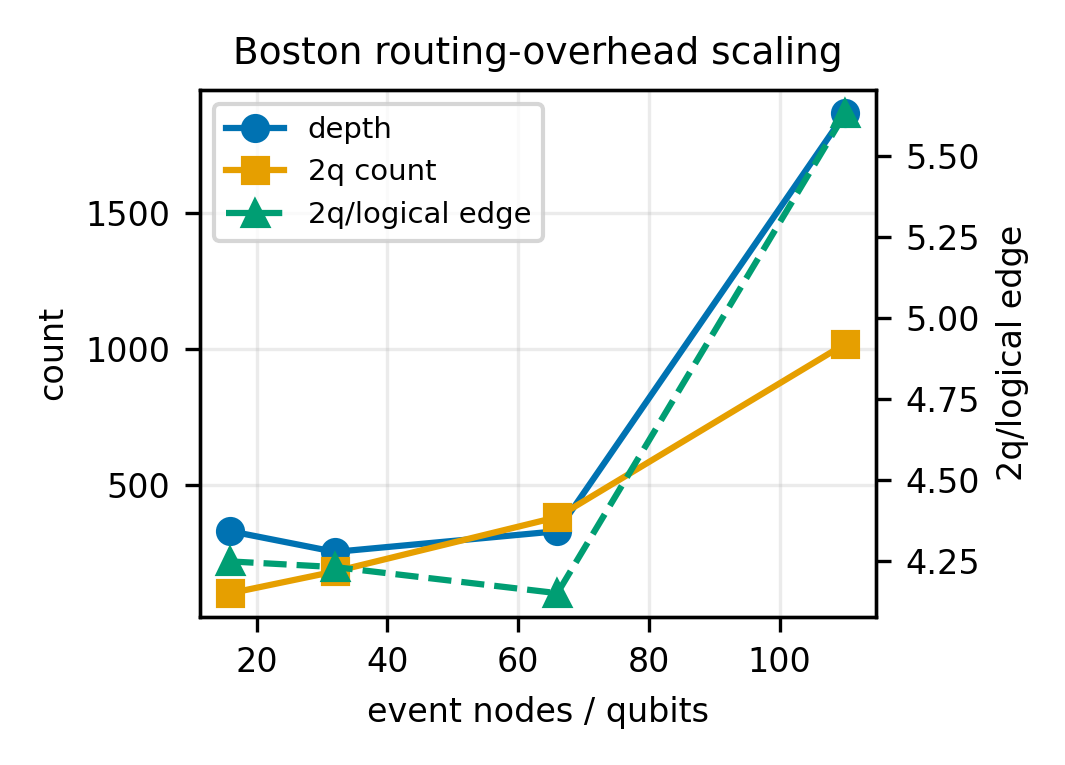}
\caption{Routing overhead scaling on \texttt{ibm\_boston}. 
Two-qubit gate count and estimated routing overhead increase substantially from H16 to H110. 
Depth is nonmonotonic because it depends on graph structure, layout, and parallelizable interactions as well as node count.}
\label{fig:boston-route-scaling}
\end{figure}

Table~\ref{tab:boston-routing-metrics} reports routing metrics on \texttt{ibm\_boston} for H16, H32, H66, and H110. Figure~\ref{fig:boston-route-scaling} plots the corresponding depth, two-qubit count, and two-qubit-per-logical-edge ratio. 
The two-qubit count and estimated SWAP-equivalent overhead increase substantially from H16 to H110. 
Depth is not strictly monotonic because it depends not only on node count, but also on graph structure, layout, routing, and the number of interactions that can be executed in parallel.

\begin{figure}[t]
\centering
\includegraphics[width=\columnwidth]{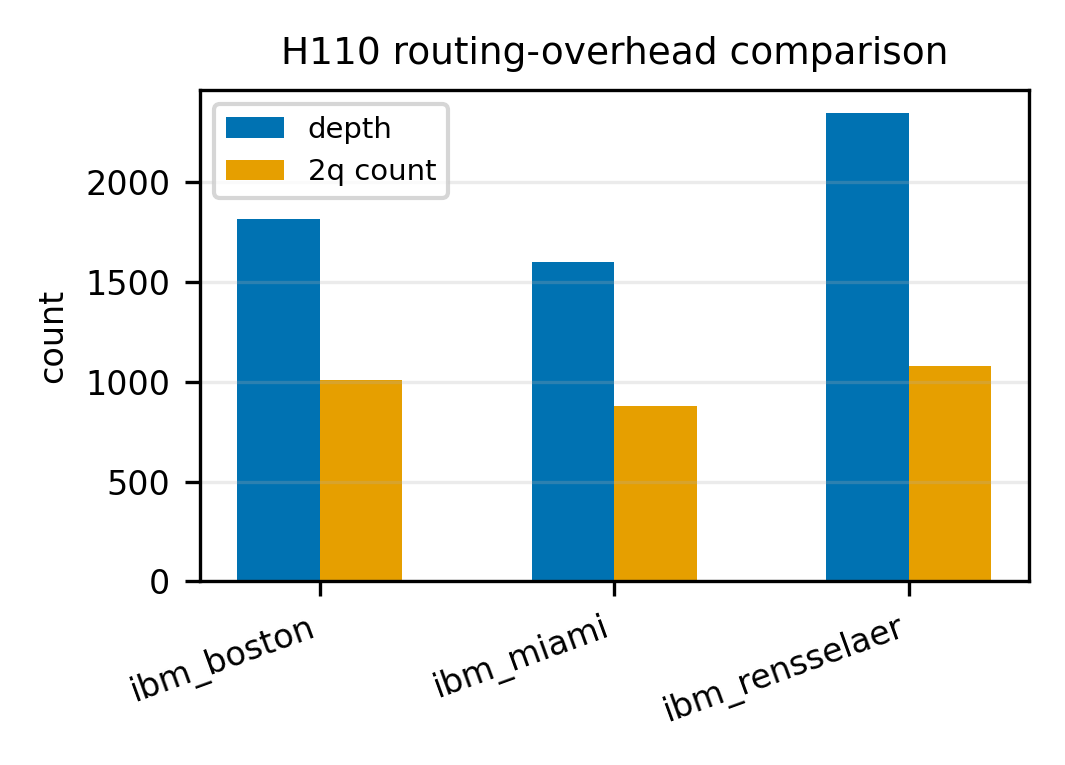}
\caption{Routing metrics for the H110 QAOA circuit across IBM backends. 
\texttt{ibm\_miami} has the lowest two-qubit overhead, \texttt{ibm\_boston} is intermediate, and \texttt{ibm\_rensselaer} has the largest overhead.}
\label{fig:h110-route-architectures}
\end{figure}

Table~\ref{tab:h110-routing-metrics} reports the H110 routing metrics across \texttt{ibm\_miami}, \texttt{ibm\_boston}, and \texttt{ibm\_rensselaer}.
Figure~\ref{fig:h110-route-architectures} visualizes these backend-dependent routing differences. 
The \texttt{ibm\_miami} transpilation has the lowest two-qubit overhead, \texttt{ibm\_boston} is intermediate, and \texttt{ibm\_rensselaer} has the largest overhead.

The routing metrics help interpret, but do not fully explain, the H110 hardware results. 
\texttt{ibm\_miami} has the lowest routing overhead, \texttt{ibm\_boston} is intermediate, and \texttt{ibm\_rensselaer} the highest, yet this ordering matches neither the objective-quality ordering nor the observed wall-clock timing. 
\texttt{ibm\_boston} attains the best mean cost ratio despite higher routing overhead than \texttt{ibm\_miami}, and \texttt{ibm\_rensselaer} records the lowest mean wall-clock time per optimizer evaluation despite the highest routing overhead. 
The latter reinforces the caveat of Section~\ref{sec:runtime-caveats} that observed wall-clock time is dominated by Runtime queue and session effects rather than by circuit complexity. 
Taken together, these results indicate that topology and routing matter, but objective quality also depends on calibration, noise, scheduling, and other backend-specific details.

\begin{table*}[t]
\centering
\caption{QAOA circuit routing metrics on \texttt{ibm\_boston}.}
\label{tab:boston-routing-metrics}
\resizebox{\textwidth}{!}{
\begin{tabular}{lrrrrrll}
\toprule
Dataset & Qubits & Logical edges & Depth & 2q count & Extra 2q & Estimated SWAP equivalents & 2q/logical edge \\
\midrule
H16 & 16 & 24 & 330 & 102 & 78 & 26.0 & 4.250 \\
H32 & 32 & 43 & 254 & 182 & 139 & 46.3 & 4.233 \\
H66 & 66 & 92 & 330 & 382 & 290 & 96.7 & 4.152 \\
H110 & 110 & 181 & 1867 & 1019 & 838 & 279.3 & 5.630 \\
\bottomrule
\end{tabular}

}
\end{table*}

\begin{table*}[t]
\centering
\caption{H110 routing metrics across IBM backends.}
\label{tab:h110-routing-metrics}
\resizebox{\textwidth}{!}{
\begin{tabular}{lrrrrll}
\toprule
Backend & Physical qubits & Coupling edges & Depth & 2q count & Estimated SWAP equivalents & 2q/logical edge \\
\midrule
\texttt{ibm\_miami} & 120 & 436 & 1597 & 878 & 232.3 & 4.851 \\
\texttt{ibm\_boston} & 156 & 352 & 1813 & 1007 & 275.3 & 5.564 \\
\texttt{ibm\_rensselaer} & 127 & 144 & 2345 & 1079 & 299.3 & 5.961 \\
\bottomrule
\end{tabular}

}
\end{table*}

\subsection{Objective Versus Security Metrics}

\begin{figure}[t]
\centering
\includegraphics[width=\columnwidth]{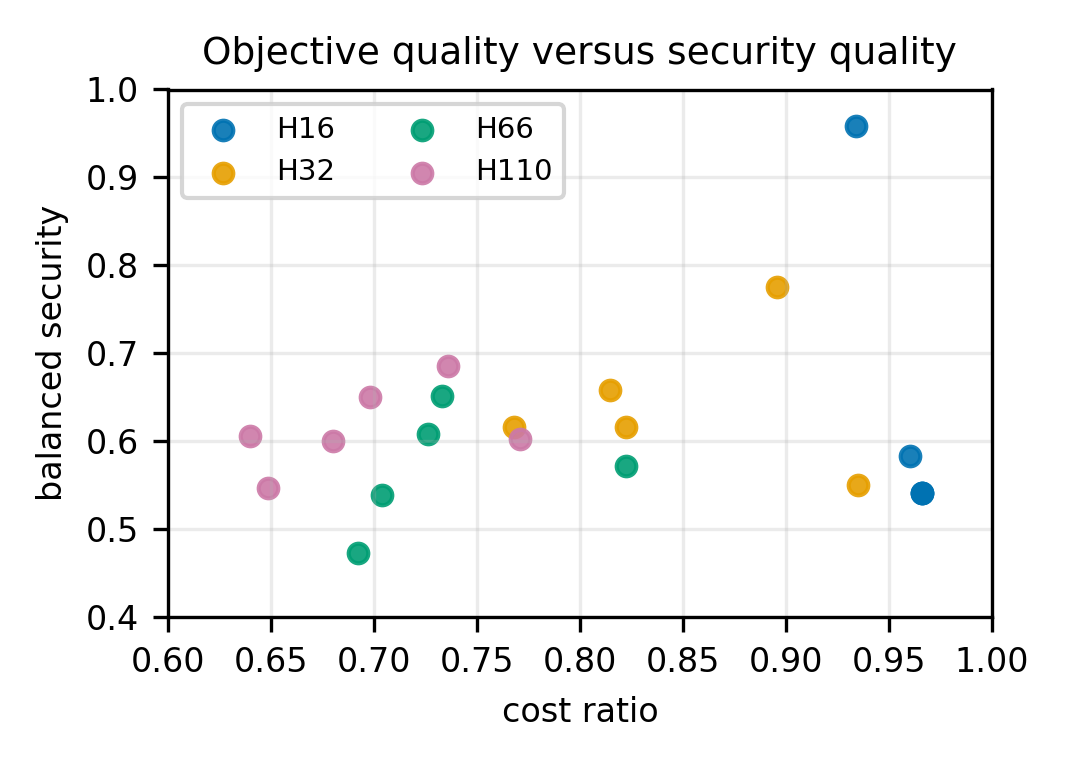}
\caption{Objective quality versus security quality for the Boston same-backend QAOA scaling runs. 
Each point is one hardware run (five seeds, 124--128, for H16, H32, and H66; six seeds, 124--129, for H110). 
MaxCut cost ratio and balanced security score are related but not equivalent: runs with similar objective quality can yield different mitigation quality, and vice versa.}
\label{fig:objective-vs-security}
\end{figure}

Figure~\ref{fig:objective-vs-security} plots MaxCut cost ratio against balanced security score for the hardware QAOA runs. 
The scatter plot shows why a single accuracy score is not sufficient for this application. 
The MaxCut cost ratio measures optimization quality relative to the class-defined reference objective, whereas balanced security measures whether the resulting partition simultaneously quarantines attack events and preserves benign events.

The relationship between these quantities is not one-to-one as runs with similar cost ratios can have different balanced security scores, and runs with lower cost ratio can sometimes have better mitigation behavior. 
This is consistent with the classical baseline results, where maximum cost cuts on H66 and H110 do not necessarily correspond to perfect attack recall and benign preservation.
Therefore, we treat MaxCut objective value as an optimization metric, and attack recall, benign preservation, quarantine precision, and balanced security as separate security metrics.

\subsection{Runtime and Measurement Caveats}
\label{sec:runtime-caveats}

The hardware runs record observed wall-clock time and optimizer time, but these quantities include IBM Runtime session behavior, job submission, result retrieval, and cloud execution variability. 
They should not be interpreted as isolated QPU execution time. As the primary indicators of circuit complexity we therefore use the routing overhead metrics: transpiled depth, two-qubit gate
count, and estimated SWAP-equivalent overhead. Observed wall-clock timing is reported for reproducibility and user-facing context, not to claim any sort of asymptotic runtime scaling.

\section{Discussion}

The present results support the proof of principle hardware claim that QAOA can be used to execute real honeypot-derived MaxCut traffic partitioning benchmark graphs on current IBM quantum hardware up to 110 event nodes. 
The study also shows that backend architecture, routing overhead, and security-specific evaluation metrics materially affect the resulting partitions.

This work should not be interpreted as a quantum advantage demonstration. 
The labeled temporal bipartite benchmark graphs studied here are structurally simple enough that greedy local search and simulated annealing reach the class-defined reference MaxCut objective. 
That result is important because it establishes the proper scope of the paper, which is to present a reproducible hardware execution and architecture benchmark for real honeypot-derived QAOA workloads.

Indeed, the value of the present work lies in the experimental framework it establishes.
The H16--H110 benchmark ladder demonstrates increasing real data graph sizes.
The H110 architecture comparison shows that different IBM processors produce different objective values, security metrics, and observed runtimes on the same 110-node QAOA workload, with the cleanest backend-level differences appearing in objective quality and routing overhead (observed wall-clock time additionally reflects IBM Runtime cloud and session effects, as discussed in Sec.~\ref{sec:runtime-caveats}).
Routing metrics connect these observations to transpilation and two-qubit routing overhead. 
Finally, the security metrics show that MaxCut cost alone is not a complete mitigation metric.

The fact that simple heuristics solve the current labeled benchmark graphs should not be interpreted as eliminating the value of the quantum experiments.
More realistic graph constructions may include noisy risk scores, source reputation, service criticality, false positive costs, time-varying traffic, or multi-action mitigation policies. 
Such formulations may produce harder QUBO instances and different tradeoffs among objective value, runtime, hardware requirements, and energy cost. 
We again note we have deliberately used a shallow QAOA ansatz and minimal optimization. 
We therefore interpret the present experiments as a first hardware feasibility and architecture study, with harder security-aware and energy-aware formulations left for future work.

Our results show the distinction between optimization quality and mitigation quality is central.
A MaxCut solution is an unordered graph partition, while a mitigation policy requires an oriented decision about which side should be quarantined. 
Moreover, when a graph has multiple connected components, different component orientations can preserve the same MaxCut objective while changing attack recall or benign preservation. 
This explains why maximum cost classical cuts on H66 and H110 do not necessarily yield a perfect balanced security. 
Future security-aware QUBO formulations should therefore encode benign preservation, false positive cost, and operational risk more directly.

\section{Limitations}

This study is intended as a proof of principle hardware feasibility and architecture benchmark experiment, and its conclusions should be interpreted in that scope. 
The benchmark graphs are constructed from labeled honeypot traces with attack and benign labels being used to form temporal bipartite graphs with known class-defined reference cuts. 
This construction is useful for controlled hardware benchmarking because it provides a transparent objective and a known security interpretation. 
However, it is not an autonomous detection method that could be used in an operational setting. 
In such a setting, labels or risk scores would need to come from an upstream honeypot rule, intrusion-detection system, firewall policy, anomaly detector, etc. 

Additionally, the graph model and edge weights are intentionally simple. 
Edges connect attack and benign events that are nearby in time, with the weights determined by temporal proximity and a fixed separation scale. 
This makes the benchmark reproducible and easy to audit, but it does not encode operational factors that may be important in practice (e.g., source reputation, traffic volume, service criticality, false positive cost, etc.). 
As a result, the current temporal bipartite graph family is not a hard classical optimization benchmark, and indeed, greedy local search and simulated annealing reach the class-defined reference MaxCut objective on all four graph instances.
Therefore, the present results are not evidence of some sort of quantum advantage. 
Their purpose, instead, is to demonstrate real hardware execution, backend-dependent behavior, routing overhead, and the need to report optimization and security metrics separately.

As mentioned, the hardware experiments also use a deliberately restricted QAOA configuration: \(p=1\), COBYLA, 1024 shots, 15 optimizer evaluations, and five seeds per condition (six for H110). 
This fixed, shallow configuration enables controlled comparisons across graph sizes and hardware backends, but it is not optimized for maximum performance. 
This framing is consistent with the role of noisy intermediate-scale quantum devices as experimental platforms for exploring near-term quantum algorithms, rather than as immediate sources of broad quantum advantage \cite{Preskill2018QuantumComputingNISQ}.
We do not make any claims that QAOA is superior to VQE or to other quantum or quantum-inspired optimizers for this application. 
A fair comparison would require a separate study of ansatz families, circuit depth, optimizer choice, initialization, shot budget, layout strategy, and hardware calibration conditions. 
Finally, the reported wall-clock times are user observed IBM Runtime measurements that include cloud execution and session effects and they should not be interpreted as isolated QPU execution times or asymptotic runtime laws. 
For this reason, we report routing overhead metrics as indicators of circuit complexity.

\section{Future Work}

The most immediate next steps involve moving from the proof or principle workflow demonstrated here towards an operational one.
Start with the graph construction.
In the present work, labels are used to construct a controlled benchmark with a known reference cut. 
Future work should instead construct graphs from observable traffic features (e.g., timestamps, source identity, source prefix, destination service, protocol, volume, session duration, geolocation, interaction type, etc.) and reveal the attack and benign labels only for evaluation. 
A closely related deployment model involves label-assisted graph optimization, in which a fast upstream detector or honeypot rule provides noisy risk scores and the graph optimizer produces a downstream mitigation decision.

A second direction is to develop security-aware QUBO objectives. 
The current MaxCut objective rewards separating weighted graph edges, but it does not directly encode information such as false positive costs, service criticality, analyst workload, or multi-action responses, among others. 
A more realistic formulation could penalize quarantining benign traffic while rewarding isolating high risk traffic, preserve critical services, or introduce additional actions (e.g., sandboxing, rate limiting, human review). 
Such formulations are likely to produce more complex (not to mention operationally meaningful) optimization graph instances than the labeled temporal bipartite graphs studied here.

Finally, the quantum hardware pipeline itself should be expanded and optimized.
Extensions include collecting additional attacks, increasing the number of hardware seeds, testing custom architecture-aware layouts, sweeping transpiler optimization levels, comparing larger \(p\) for a more expressive QAOA circuit, alternating optimizers, investigating VQE ansatz families, and using bespoke and/or commercial QUBO optimizers. 
A longer term goal is to evaluate not only objective quality and security quality, but the total cost of solution, including QPU time, classical optimizer time, classical baseline energy, and end-to-end mitigation latency.

\section{Conclusion}
In this work we presented a reproducible pipeline for mapping real honeypot event traces to weighted MaxCut benchmark graphs and executing the resulting QAOA circuits on IBM quantum hardware. 
The experimental dataset comprises four labeled temporal bipartite benchmark instances (H16, H32, H66, H110), a single backend scaling ladder on \texttt{ibm\_boston}, a matched H110 architecture comparison across the three backends \texttt{ibm\_rensselaer}, \texttt{ibm\_miami}, and \texttt{ibm\_boston}, classical heuristic baselines, and a routing overhead audit. 
The results demonstrate that a shallow QAOA can execute 100+ qubit cybersecurity-derived MaxCut problems on current hardware, that backend architecture affects objective quality, security quality, and routing overhead even when the logical graph and algorithmic settings are held fixed, and that MaxCut cost ratio, attack recall, benign preservation, quarantine precision, and balanced security must be reported as separate metrics. 
Because simple classical heuristics solve the present labeled benchmark graphs, this work is a hardware feasibility and architecture benchmark rather than a quantum advantage claim, and motivates label-free graph construction, security-aware QUBO objectives, and energy-aware cost of solution studies as next steps.

\section*{Data and Code Availability}
Data and code are available at \url{https://github.com/cogburner/qddos-honeypot-qaoa}.

\section*{Acknowledgments}
CVC is supported in part by the RPI-IBM Future of Computing Research Collaboration. 

\balance
\bibliographystyle{IEEEtran}
\bibliography{references}
\end{document}